\documentclass[reprint,
superscriptaddress, 
amsmath,amssymb,
aps, 
pra, 
10pt, 
notitlepage, 
nofootinbib,
tightenlines]
{revtex4-2}

\usepackage[export]{adjustbox}
\usepackage{amsfonts}
\usepackage{amsmath}
\usepackage{amssymb}
\usepackage{amsthm}
\usepackage[title]{appendix}
\usepackage{changepage}
\usepackage{dsfont}
\usepackage[colorlinks=True, citecolor=blue, urlcolor=blue]{hyperref}
\usepackage[utf8]{inputenc}
\usepackage{xcolor}
\usepackage{wisjab}

\newtheorem{theorem}{Theorem}

\def\a{\mt a}
\def\ata#1#2{\a_{#1}\t\a_{#2}}
\def\atataa#1#2#3#4{\a_{#1}\t\a_{#2}\t\a_{#3}\a_{#4}}

\def\b{\mt b}
\def\n{\mt n}
\def\L{L}
\def\P{\mathcal P}
\def\U{\mathcal U}

\def\up{\uparrow}
\def\down{\downarrow}

\def\H{\mt H}
\def\Hi{\mt H^{\rm i}}
\def\Hf{\mt H^{\rm f}}
\def\Hr{\mt H^{\rm r}}

\let\mtgXi\Xi

\let\mtgSigma\Sigma

\let\mtgPhi\Phi

\let\Gamma\varGamma
\let\Delta\varDelta
\let\Theta\varTheta
\let\Lambda\varLambda
\let\Xi\varXi
\let\Pi\varPi
\let\Sigma\varSigma
\let\Upsilon\varUpsilon
\let\Phi\varPhi
\let\Psi\varPsi
\let\Omega\varOmega

\begin{document}

\title{Adiabatic ground state preparation of fermionic many-body systems from a two-body perspective}

\author{Dyon van Vreumingen}
\affiliation{Institute for Theoretical Physics, Universiteit van Amsterdam, The Netherlands}
\affiliation{QuSoft, Centrum Wiskunde en Informatica, The Netherlands}
\author{Kareljan Schoutens}
\affiliation{Institute for Theoretical Physics, Universiteit van Amsterdam, The Netherlands}
\affiliation{QuSoft, Centrum Wiskunde en Informatica, The Netherlands}

\date{\today}

\begin{abstract}
A well-known method to prepare ground states of fermionic many-body hamiltonians is adiabatic state preparation, in which an easy to prepare state is time-evolved towards an approximate ground state under a specific time-dependent hamiltonian. However, which path to take in the evolution is often unclear, and a direct linear interpolation, which is the most common method, may not be optimal. In this work, we explore new types of adiabatic paths based on the spectral decomposition of the two-body projection of the residual hamiltonian (the difference between the final and initial hamiltonian). The decomposition defines a set of hamiltonian terms which may be adiabatically interpolated in a piecewise or combined fashion. 
We demonstrate the usefulness of partially piecewise interpolation through examples involving Fermi-Hubbard models where, due to symmetries, level crossings occur in direct (fully combined) interpolation. We show that this specific deviation from a direct path appropriately breaks the relevant symmetries, thus avoiding level crossings and enabling an adiabatic passage.
On the other hand, we show that a fully piecewise scheme, which interpolates every hamiltonian term separately, exhibits a worst-case complexity of $O(L^6/\Delta^3)$ as compared to $O(L^4/\Delta^3)$ for direct interpolation, in terms of the number of one-body modes $L$ and the minimal gap $\Delta$ along the path. This suboptimality result suggests that only those terms which break necessary symmetries should be taken into account for piecewise interpolation, while the rest is treated with direct interpolation.
\end{abstract}

\maketitle

\section{Introduction}
Quantum computers are currently regarded as a prime candidate for solving problems in condensed matter physics and chemistry that are untractable for classical computers. In particular, since Feynman's observation of the potential of quantum simulation \cite{Feynman1982}, the pioneering work by Lloyd \cite{Lloyd1996} and the invention of quantum phase estimation \cite{Kitaev1995}, interest in the deployment of quantum computers as simulators of highly correlated quantum systems has exploded.
A large body of work has been established describing techniques for simulating dynamics of many-body systems on a quantum computer \cite{Whitfield2011, Low2016, Babbush2018a, Babbush2018b, Kivlichan2020, Lee2021}, and these may be combined with quantum phase estimation in order to estimate eigenenergies \cite{VonBurg2021}. A critical question however, to make these methods useful, is how to prepare the states of interest -- be it thermal states or eigenstates of the system under investigation -- that serve as input to the algorithms that simulate dynamics or compute energies. Although experimental efforts using heuristics such as variational quantum eigensolvers \cite{Cade2019, Montanaro2020, Wei2020, Tilly2022} have shown great success in preparing such states for systems of fixed size, much remains unknown with regards to ``solving'' highly correlated systems in general.\par
A well-known method for preparing approximate ground states of complex systems is adiabatic state preparation, which uses the adiabatic theorem to carry out quantum computation. While originally formulated as a tool to approximate quantum dynamics on large time scales with respect to the inverse energy gap, the adiabatic theorem was reintroduced to attack combinatorial problems \cite{Farhi2000} and to study many-body systems such as Fermi-Hubbard models \cite{Wecker2015, Perez2022} and molecules \cite{Du2010, Veis2014, Babbush2014, Sugisaki2022}.\par
The idea of adiabatic state preparation (ASP) is to prepare an eigenstate $\ket{\psi^{\rm f}}$ of a ``final'' hamiltonian $\Hf$, starting with an eigenstate $\ket{\psi^{\rm i}}$ of an ``initial'' hamiltonian $\Hi$ which is straightforward to prepare. Given this initial state, one time-evolves the state according to the time-rescaled Schr\"odinger equation,
\begin{align}\label{eq:sTDSE}
i\af{}s\ket{\psi(s)} = T\,\mt H(s)\ket{\psi(s)},
\end{align}
where $s = t/T$ is a dimensionless time, and $T$ is the total (physical) evolution time. (We work in units such that $\hbar = 1$.) The evolution is carried out under a time-dependent hamiltonian $\mt H(s)$ which equals $\Hi$ at $s=0$ and $\Hf$ at $s=1$. After an evolution with time $s$, one obtains a state $\ket{\psi^T(s)} = \mt U^T(s)\ket{\psi(0)}$, where $\ket{\psi(0)} = \ket{\psi^{\rm i}}$ and $\mt U^T(s)$ solves eq.~\ref{eq:sTDSE}. By what is known as the \emph{adiabatic theorem}, the state at the end of this evolution, $\ket{\psi^T(1)}$ will be close to the final eigenstate $\ket{\psi^{\rm f}}$ if $T$ is sufficiently large. One variant of this adiabatic theorem which precisely indicates what ``close'' and ``sufficiently large'' mean in this context, is due to Jansen et al. \cite{Jansen2006}. The statement is that if $\H(s)$ is a hamiltonian defined on the interval $[0, 1]$ which for every $s\in[0, 1]$ has an instantaneous eigenstate $\ket{\psi(s)}$ whose energy is separated from the rest of the spectrum by $\Delta(s) > 0$, then for any $s\in[0, 1]$, the condition
\begin{align}\label{eq:JRS_T}
    T \geq \frac1\delta \bigg( \int_0^s \bigg[ \frac{\|\partial_s^2{\mt H}(\sigma)\|}{\Delta^2(s)} + 7\frac{\|\partial_s{\mt H}(\sigma)\|^2}{\Delta^3(s)} \bigg] \, d\sigma + B \bigg)
\end{align}
where $\|\cdot\|$ denotes the operator norm 
and $B$ is a boundary term that may be set to zero if $\dot{\mt H}(0) = \dot{\mt H}(1) = 0$, is sufficient to guarantee that
\begin{align}
    |\braket{\psi(s)}{\psi^T(s)}| \geq 1 - \delta
\end{align}
provided that $\ket{\psi^T(0)} = \ket{\psi(0)}$.
Throughout the rest of this paper, we will consider the case where $\ket{\psi(s)}$ is the ground state of $\mt H(s)$.\par
In principle, any adiabatic evolution may be implemented on a gate-based quantum computer through Trotter-Suzuki \cite{Suzuki1993, Childs2019} or more sophisticated time-dependent hamiltonian simulation methods \cite{Wan2022, Low2019, Kieferova2018}. Alternative approaches approximate the evolution through a series of measurements \cite{Aharonov2003, Lemieux2021} or simulations thereof \cite{Boixo2009, Boixo2010}.\par 
The most commonly used interpolation method in adiabatic state preparation is a direct linear interpolation between $\Hi$ and $\Hf$ \cite{Albash2016}, which is to say that
\begin{align}\label{eq:directinterp}
    \mt H(s) = \Hi + s(\Hf - \Hi).
\end{align}
However, this method is rather restrictive as the evolution is controlled by only a single parameter, $s$. Thus the evolution is sensitive to gap closures along the path, which cannot be avoided. An obvious solution is to increase the number of control parameters in the passage from $\Hi$ to $\Hf$. This approach is discussed by Tomka et al. \cite{Tomka2016}, who show that evolving along a geodesic path, based on the quantum metric tensor (or Fubini-Study metric) with respect to the control parameters, maximises the local fidelity along the path. In addition, they show that an increase in the number of control parameters leads to higher final fidelities. Put simply, their results rely on the fact that geodesic paths ``walk around'' regions of parameter space associated with small energy gaps, thus minimising diabatic errors.
A similar category of methods to avoid problematic regions in adiabatic state preparation is known as counteradiabatic driving, where an additional hamiltonian term is added during the evolution, which actively suppresses diabatic errors and is set to zero at the end \cite{Demirplak2003, Demirplak2005, Demirplak2008}.\par
The problem with these approaches, however, is that their implementation becomes infeasible for large, complex systems. For counterdiabatic driving to work, the eigenstates and spectrum of the hamiltonian must be known along the path, which is something we cannot expect to achieve in such settings. For the geodesic approach, the main roadblock is the inability to solve the geodesic equations, which become inaccesibly large systems of differential equations already for small many-body problems.\par
In this work, we introduce a more hands-on approach to produce new types of adiabatic paths for generic fermionic many-body hamiltonians in a second quantised representation. Section \ref{sec:hamils} gives a brief description of such systems. The adiabatic paths are based on a decomposition of the coefficient tensor of such hamiltonians (section \ref{sec:idee}), which defines a set of control parameters that govern the adiabatic evolution. We emphasise that such adiabatic paths can be seen as a new view on adiabatic state preparation for fermionic systems by considering many-body hamiltonians in terms of their two-body eigenstates. 
We demonstrate, through a set of worked examples (section \ref{sec:vb}), that there exist scenarios in which direct interpolation suffers from level crossings caused by symmetries, and how the two-body decomposition may be used to explicitly break symmetries and lift such crossings. 
In section \ref{sec:opnorm}, we show how a description of these two-body eigenstates as superpositions of fermion pairs, following a suitable one-body transformation, leads to a worst-case adiabatic complexity in terms of the number of one-body modes $L$ and a minimum gap $\Delta$ (section \ref{sec:opnorm}). The implications of this analysis are discussed for different systems.
We summarise and conclude in section \ref{sec:discussion}.

\section{Many-body hamiltonians \label{sec:hamils}}
Of interest in this work are generic fermionic, interacting, particle-conserving many-body hamiltonians, expressed in a second-quantised representation as
\begin{align}\label{eq:hamiltonian}
\H = &\sum_{P, Q = 1}^{\L} h_{PQ}\, \ata PQ \nonumber\\
&+ \frac12 \sum_{P, Q, R, S = 1}^{\L} g_{PQRS}\, \atataa PRSQ
\end{align}
where $P, Q, R, S$ index general single-particle modes (which may include a spin index, in which case the modes are known as spin orbitals). Furthermore, the coefficients $h_{PQ}$ and $g_{PQRS}$ describe the one- and two-body terms respectively, and the fermionic creation (annihilation) operators $\a_P\t$ ($\a_P$) satisfy the canonical anticommutation relations,
\begin{align}
    &\{\a_P, \a_Q\} = \{\a_P\t, \a_Q\t\} = 0, \nonumber\\
    &\{\a_P\t, \a_Q\} = \delta_{PQ}.
\end{align}
Depending on context, we will sometimes split a single-particle mode into a spatial and a spin component, writing lowercase $p, q, r, s$ for the spatial and $\sigma, \upsilon, \tau, \varphi$ for the spin component.\par
Such hamiltonians are the central object of study in chemistry and condensed matter theory. In chemistry, the starting point for describing molecules is typically the electronic structure hamiltonian in the nonrelativistic Born-Oppenheimer approximation, given in first quantisation by
\begin{align}
    \hat{\H} 
    = E_{\rm nuc}\underbrace{-\, \sum_{Ii}\frac1{|\vc r_I - \vc r_i|} + \frac12 \sum_i \nabla_i^2}_{\hat{\mt h}} + \underbrace{\frac12 \sum_{ij} \frac1{|\vc r_i - \vc r_j|}}_{\hat{\mt g}}
\end{align}
where the upper case indices run over all nuclei and the lower case indices label the electrons. $E_{\rm nuc}$ is a nuclear energy constant which may be set to zero for practical purposes. 
When we project the Hilbert space onto a fixed basis set $\{\ket{\psi_{p\sigma}}\}$ consisting of single-particle modes (which may be assumed to be real, following common practice), the projected electronic structure hamiltonian assumes the form of eq.~\ref{eq:hamiltonian} with
\begin{align}
    h_{(p\sigma)(q\upsilon)} &= h_{pq}\delta_{\sigma\upsilon}, ~~ g_{(p\sigma)(q\upsilon)(r\tau)(s\varphi)} = g_{pqrs}\delta_{\sigma\upsilon}\delta_{\tau\varphi}, \label{eq:h_pq,g_pqrs,spindelta}\\[2mm]
    h_{pq} &= \int d\vc r_1\,\psi_p(\vc r_1) \,\hat{\mt h}\, \psi_q(\vc r_1), \\
    g_{pqrs} &= \int d\vc r_1d\vc r_2\,\psi_p(\vc r_1)\psi_q(\vc r_1) \,\hat{\mt g}\, \psi_r(\vc r_2)\psi_s(\vc r_2).
\end{align}
From these expressions we may draw the symmetry conditions
\begin{align}
    h_{PQ} &= h_{QP}\label{eq:symmetries_h} \\
    g_{PQRS} = g_{RSPQ} &= g_{QPRS} = g_{PQSR}\label{eq:symmetries_g}
\end{align}
which we shall assume satisfied for all hamiltonians considered in this paper. Note that the hermiticity of the hamiltonian is guaranteed by the use of real-valued single-particle modes.\par
The electronic structure hamiltonian may be simplified by restricting the electrons to orbitals localised at sites arranged on a lattice, and neglecting any Coulomb interaction between different sites. Taking one orbital per site, one arrives at the single-band Fermi-Hubbard (FH) hamiltonian,
\begin{align}\label{eq:fermi-hubbard}
    \H = j \sum_{\g{p, q}, \sigma} (\ata{p\sigma}{q\sigma} + \ata{q\sigma}{p\sigma}) + U \sum_{p}\mt n_{p\uparrow}\mt n_{p\downarrow} + \mu\sum_{p\sigma}\n_{p\sigma}
\end{align}
where $j$ is the hopping strength between two neighbouring sites, $U$ is the on-site Coulomb interaction and $\mu$ is a chemical potential strength. The Fermi-Hubbard model may be written in the form of eq.~\ref{eq:hamiltonian} with coefficients as in eq.~\ref{eq:h_pq,g_pqrs,spindelta} through the identification
\begin{align}
    h_{pq} &= \begin{cases}
    j & \text{if sites $p$ and $q$ are neighbours},\\
    0  & \text{else},
    \end{cases}\nonumber\\[1mm]
    g_{pqrs} &= U\delta_{pq}\delta_{rs}\delta_{pr}
\end{align}
which are readily seen to exhibit the symmetries of eqs.~\ref{eq:symmetries_h}--\ref{eq:symmetries_g}. In section \ref{sec:vb}, we study a generalisation of the one-dimensional FH hamiltonian that allows for spin-dependent hopping strengths, i.e. $h_{(p\sigma)(q\upsilon)} = h^\sigma_{pq}\delta_{\sigma\upsilon}$.\par
Throughout the rest of this paper, we will work with a fixed particle number (denoted $N$) for each hamiltonian. In a sector of fixed $N$, we can absorb the one-body terms of any hamiltonian of the form in eq.~\ref{eq:hamiltonian} into the two-body terms, by inserting the identity as 
\begin{align}
\ata PQ = \frac1{N-1}\sum_R\a_P\t\a_R\t\a_R\a_Q \label{eq:aPtaQidentity}
\end{align}
and when we define
\begin{align}\label{eq:w_pqrs}
w_{PQRS} := \frac{h_{PQ}\delta_{RS} + \delta_{PQ}h_{RS}}{N-1}
\end{align}
then we may write the one-body operator as
\begin{align}
\sum_{PQ} h_{PQ}\, \ata PQ = \frac12\sum_{PQRS} w_{PQRS} \,\atataa PRSQ.
\end{align}
Now define the combined one-body and two-body interaction tensor
\begin{align}
G_{PQRS} := \frac12(w_{PQRS} + g_{PQRS})
\end{align}
and observe that
\begin{align}
\mt H = \sum_{PQRS} G_{PQRS}\, \atataa PRSQ;
\end{align}
the one- and two-electron terms have now been combined into a single term. We shall refer to $\mt G$ as the interaction tensor of $\mt H$.\par
Lastly, it will be convenient to express $\mt H$ as a sum over only the unique pairs $(P, R)$ and $(Q, S)$: by invoking the fermionic anticommutation relations, we may write
\begin{align}\label{eq:generic_H_antisymmetrised}
\mt H = \sum_{\substack{P<R\\Q<S}} \tilde G_{PQRS}\, \atataa PRSQ
\end{align}
with $\tilde G_{PQRS} := G_{PQRS} - G_{PSRQ} - G_{RQPS} + G_{RSPQ} = 2(G_{PQRS} - G_{PSRQ})$. This tensor shall be termed the antisymmetrised interaction tensor of $\mt H$.\par
Naturally, if the initial hamiltonian $\Hi$ contains all one-body terms, then there is no need to explicitly include them in the two-body terms of the residual hamiltonian $\Hr$ as described above. In section \ref{sec:vb} we will see an example of this where $\Hi$ takes the form of a mean-field (also known as Hartree-Fock) approximation.

\section{Adiabatic state preparation by two-body eigendecomposition \label{sec:idee}}
In this work, we deviate from the direct interpolation approach (eq.~\ref{eq:directinterp}) by decomposing the residual hamiltonian $\Hr = \Hf - \Hi$ into a sum of terms $\Hr_k$, $k\in\{1, \ldots, M\}$, and evolving a linear combination of the terms $\Hr_k$. That is,
\begin{align}\label{eq:newpath}
    \mt H(s) = \Hi + \sum_{k=1}^M \gamma_k(s) \, \Hr_k.    
\end{align}
The decomposition defines an $M$-dimensional parameter space in which we consider paths $\vcg\gamma(s)$ restricted to a hypercube, starting from $\vcg\gamma(0) = [0, 0, \ldots, 0]$ and ending at $\vcg\gamma(1) = [1, 1, \ldots, 1]$.\par
We define the residual terms $\Hr_k$ through a decomposition of the antisymmetrised interaction tensor of $\Hr$. This is akin to known low-rank factorisation methods of the two-body part of the interaction tensor, aimed at achieving improved memory efficiency and speed-ups in quantum simulation implementations \cite{Rubin2022, Lee2021}. Whereas most of these works focus on the Cholesky decomposition of the interaction tensor \cite{Koch2003, Nottoli2021, Roeggen2008}, we use an eigendecomposition. More precisely, we regard the antisymmetrised interaction tensor $\tilde{\mt G}$ as an $\L(\L-1)/2 \times \L(\L-1)/2$ matrix $\mt F$, by combining the indices $(PR)$ and $(QS)$:
\begin{align}
F_{(PR)(QS)} := \tilde G_{PQRS}. \label{eq:Fdef}
\end{align}
We point out that $\mt F$ represents a two-particle hamiltonian, since in the two-particle sector, the operator string $\atataa PRSQ$ is equivalent to the outer product $\ket{PR}\bra{QS}$. For this reason, we refer to $\mt F$ as the two-particle matrix. Note however that the noninteracting part of $\mt F$ is scaled by a factor $1/(N - 1)$ with respect to that of $\Hr$, which arises from the insertion of identity (eq.~\ref{eq:aPtaQidentity}).\par
Now since $\mt F$ is symmetric with respect to the exchange $(PR)\leftrightarrow(QS)$ (following from the symmetry conditions of eq.~\ref{eq:symmetries_h}--\ref{eq:symmetries_g}, we may eigendecompose $\mt F$ into (normalised) orthogonal eigenvectors,
\begin{align}
F_{(PR)(QS)} = \sum_k \lambda_k \, \phi^{(PR)}_k \phi^{(QS)}_k   
\end{align}
and define
\begin{align}\label{eq:nqf_1st_decomposition}
\Hr_k := \lambda_k \Big( \sum_{P<R} \phi^{(PR)}_k \a_P\t \a_R\t \Big) \Big( \sum_{Q<S} \phi^{(QS)}_k \a_S \a_Q \Big) =: \lambda_k \mtgPhi_k.
\end{align}
In the following, we will refer to $\lambda_k$ as the two-body eigenvalues; the states $\ket{\phi_k} = \sum_{P<R} \phi^{(PR)} \a_P\t\a_R\t \ket{\rm vac}$ as the two-body eigenstates; and the operators $\mtgPhi_k$ as the pseudoprojectors of $\mt F$.

\section{Lifting crossings by symmetry breaking in Fermi-Hubbard models \label{sec:vb}}
We will now illustrate how our method applies to cases where a discrete symmetry in the hamiltonian is influential on the course of the adiabatic evolution. In particular, we consider the situation in which the initial hamiltonian and the final hamiltonian share such a symmetry. In such a case, if the ground states of the initial and final hamiltonian belong to different symmetry sectors, then necessarily at some point the energy levels cross and an excited state is obtained at the end of the adiabatic evolution. This is a known problem that was addressed in the work by Farhi et al. \cite{Farhi2000} and also plays a role in many-body contexts \cite{Francis2022}. Typically the solution is to add a symmetry-breaking field to the interpolation hamiltonian which is set to zero in the end. In this section, we show how symmetry breaking behaviour emerges naturally from the formalism of the two-body eigendecomposition.\par
To see where this symmetry breaking comes from, it is important to note the following fact: if and only if a two-particle eigenstate $\ket\phi = \sum_{P<R} \phi_{(PR)} \a_P\t\a_R\t\ket{\rm vac}$ is also an eigenstate of some unitary symmetry operator $\mt U$ (with some eigenvalue $\mu$), which is expressible as a product of one-body rotations, then the corresponding two-body operator $\mtgPhi$ commutes with $\mt U$. The ``only if'' direction is trivial (since $\ket\phi$ is an eigenstate of $\mtgPhi$); for the ``if'' direction, observe that
\begin{align}
    \mu\ket\phi &= \mt U \Big( \sum_{P<R} \phi_{(PR)} \, \a_P\t\a_R\t \ket{\rm vac} \Big) \nonumber\\
    &= \sum_{P<R} \phi_{(PR)} \mt U\a_P\t\mt U\t \mt U\a_R\t\mt U\t \ket{\rm vac} \nonumber\\
    &= \sum_{P<R} \phi_{(PR)} \, \Big( \sum_M U_{PM} \a_M\t \Big) \Big( \sum_N U_{RN}\a_N\t \Big) \ket{\rm vac} \nonumber\\
    &= \sum_{M<N} \overbrace{ \sum_{P<R} \phi_{(PR)} \big[ U_{PM} U_{RN} - U_{PN} U_{RM} \big] }^{= \; \mu \, \phi_{(MN)}} \times \nonumber\\[-1mm]
    & \hspace*{2cm} \times \a_M\t\a_N\t \ket{\rm vac}.
\end{align}
The last line and the fact that $|\mu| = 1$ then imply that
\begin{align}
    \mt U \mtgPhi \mt U\t = \mu \mtgPhi \mu^* = \mtgPhi
\end{align}
so that indeed $[\mtgPhi, \mt U] = 0$. This has the following implication: if all two-body eigenvalues $\lambda_k$ are distinct, then all two-particle states $\ket{\phi_k}$, being eigenstates of the residual hamiltonian $\Hr$, are also eigenstates of the symmetry operator $\mt U$, and thus all hamiltonian terms $\Hr_k$ commute with $\mt U$. In such a case, no symmetry is broken. However, if the eigenvalues corresponding to two two-body eigenstates with different symmetry are degenerate, then these two-body eigenstates may be mixed to produce new hamiltonian terms which do not commute with $\mt U$ and therefore break the symmetry. In practice, one will need to fix a particular mixing to make the adiabatic process unambiguous; in the following, this is taken care of by a small splitting $d\lambda$ in the relevant two-body eigenvalues. Note that while such a splitting does open a gap, this gap scales inversely in $d\lambda$; thus if this splitting is small with respect to the overall energy scale, direct linear interpolation still requires a problematically large evolution time. \par
We demonstrate this idea with two simple examples, chosen such that (i) it displays an approximate discrete symmetry, which is only slightly broken and, (ii) the best mean field (Hartree-Fock) solution predicts a ground state in a symmetry sector different from that of the true ground state. In such a situation, straightforward adiabatic following of the Hartree-Fock state has to be exceedingly slow to avoid a crossing into an excited state. A multi-step adiabatic procedure, along the lines presented in this paper, will avoid the crossing altogether and allow a convergence on the true ground state.
\subsection{Fermi-Hubbard trimer}
First, consider the following two-particle, three-site Fermi-Hubbard model:
\begin{align*}
\mt H = \includegraphics[width=0.215\textwidth, valign=c]{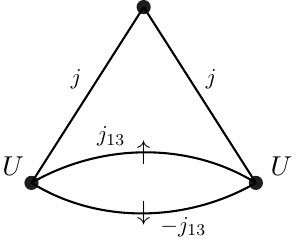}
\end{align*}
\begin{align}
= {}& \sum_\sigma j(\ata{1\sigma}{2\sigma} + \ata{3\sigma}{1\sigma}) + j_\sigma\, \ata{1\sigma}{3\sigma} + \text{h.c.} \nonumber\\
&{} + U (\n_{1\uparrow}\n_{1\downarrow} + \n_{3\uparrow}\n_{3\downarrow})
\end{align}
where $\n_{i\sigma} = \ata{i\sigma}{i\sigma}$; $h$, $j_\uparrow$, $j_\downarrow$ and $U\leq0$ are real constants; and we set $j_\uparrow = j_{13} = -j_\downarrow$. The discrete symmetry we will keep track of is the reflection of sites $1 \leftrightarrow 3$. It is exact when $U_1=U_3$, and we will later consider cases where $U_1$ and $U_3$ are slightly different, breaking the symmetry. We assume $j>0$ throughout.
\subsubsection{The case $U = 0$}
Let us first consider $U_1=U_3=0$. For $j_13=j$, the one-body kinetic energy terms for spin up have two degenerate ground states, at energy $E_{{\rm kin}\up}=-j$. For $j_{13}<j$, the unique ground state is symmetric (S), while for $j_{13}>j$ the ground state is anti-symmetric (A). This makes clear that the ground state for one spin-up and one spin-down particle is symmetric for $j_{13}<j$ but antisymmetric for $j_{13}>j$. Turning on $U<0$ will shift the S-A transition to lower values of $j_{13}$. The Hartree-Fock mean field solution follows this trend but we will see that there are values of $j_{13}$ where Hartree-Fock places the ground state in the wrong symmetry sector.
\subsubsection{The case $U<0$: tracing the ground state of $\mt H$}

Turning on a negative $U_1=U_3=U$ will change the nature of the two-body ground state. 

For small $|U| \ll j$ and $j_{13}=j+\delta$ the energies of the symmetric (S) and anti-symmetric (A) states split as (in first order perturbation theory in $U$, $\delta$)
\begin{align}
    E_{\rm S}= -3j + U/9 - \delta/3, ~~ E_{\rm A}= -3j + U/3 - 5 \delta/3
\end{align}
implying that the S-A crossing (as a function of $j_{13}$) shifts to $j_{13}=j+U/6$, that is to a smaller value of $j_{13}$.\par
For $U$ large and negative, the two electrons will tend to form a local pair at site 1 or 3, with energy $U$. In second order perturbation theory, taking into account processes with two hops (of strength $j$ or $j_{13}$) connecting the pair states with unpaired states at energy 0,  the on-site energies of these pairs are adjusted to
\begin{align}
\epsilon_1=\epsilon_3=U + 2 j^2/U + 2 j_{13}^2/U
\end{align}
while the pair hopping amplitude becomes
\begin{align}
t_{13}= -2 j_{13}^2/U.
\end{align}
This leads to S and A ground state energies 
\begin{align}
E^{(2)}_{\rm S}=U+2 j^2/U, ~~~~ E^{(2)}_{\rm A}=U+2 j^2/U+4 j_3^2/U.
\end{align} 
Including terms of order $j^4/U^3$, $j^2 j_{13}^2/U^3$ and $j_{13}^4/U^3$ we find
\begin{align}
E^{(4)}_{\rm S} &= U+ 2 {j^2 \over U} + 8 {j^4 \over U^3} + {14 \over 3} {j^2 j_{13}^2 \over U^3} + O\Big(\frac{(j^2 + j_{13}^2)^3}{U^5}\Big),
\\[2mm]
E^{(4)}_{\rm A} &= U + 2 {j^2 \over U} + 4 {j_{13}^2 \over U} - 4 {j^4 \over U^3}  + 2 {j^2 j_{13}^2 \over U^3} - 16 {j_{13}^4 \over U^3} \nonumber\\ & ~~~~ + O\Big(\frac{(j^2 + j_{13}^2)^3}{U^5}\Big).
\end{align}
This puts the S-A crossing (in an expansion in terms of $j/U$) at $j_{13} = \sqrt{3}j^2/|U|$.

\subsubsection{The case $U<0$: Hartree-Fock approximation}
To write a mean field (Hartree-Fock) Ansatz, we should first decide on the symmetry sector. For an overall antisymmetric Ansatz, we have
\begin{align}
\ket{\rm HF, A} = {1 \over \sqrt{2}} (\a\t_{1\up} - \a\t_{3\up}) {1 \over \sqrt{2+x^2}} (\a\t_{1\down}+x \a\t_{2\down}+\a\t_{3\down}) \ket{\rm vac} .
\end{align}
The expectation value becomes
\begin{align}
\langle \mt H \rangle_{\rm HF, A}
&= {1 \over 2+x^2} \left( 4 j x - 2 j_{13}  - (2+x^2) j_{13} + U \right) \nonumber\\
&= {1 \over 2+x^2} \left( (U-4 j_{13})+ 4j x - j_{13} x^2 \right).
\end{align}
This expression is minimised for (keeping the leading terms in an expansion in terms of $j/U$, $j_{13}/U$)
\begin{align}
& x=4j/U ~~ \Rightarrow \nonumber\\[1mm]
\langle \H \rangle_{\rm HF, A}^{\rm min} = {U \over 2} - 2 &j_{13} + 4 {j^2 \over U}+ O\Big({(j + j_{13})^3 \over U^2}\Big).
\end{align}
The competing Ansatz is symmetric in both the up and the down factors, 
\begin{align}
\ket{\rm HF, S} &= {1 \over \sqrt{2+y^2}} (\a\t_{1\up}+ y \a\t_{2\up} + \a\t_{3\down}) \times {}\nonumber\\
& ~~~~ \times {1 \over \sqrt{2+x^2}} (\a\t_{1\down}+x \a\t_{2\down}+\a\t_{3\down}) \ket{\rm vac} ,
\end{align}
leading to
\begin{align}
\langle \mt H \rangle_{\rm HF, S}
&= {1 \over (2+x^2)(2+y^2)} \big[ 4 j x (2+y^2) + 4 j y (2+x^2) \nonumber\\
& ~~~~ - 2 (2+y^2) j_{13} + 2 (2+x^2) j_{13} + 2U \big] \nonumber \\[1mm]
&= {1 \over (2+x^2)(2+y^2)} \big[ 8 j (x+y) + 4 j (x y^2+y x^2) \nonumber\\
& ~~~~ + 2 (x^2-y^2) j_{13} + 2U \big] .
\end{align}
In leading order, the minimum energy is reached for
\begin{align}
&x=y=4j/U ~~ \Rightarrow \nonumber\\[1mm]
\langle \mt H \rangle_{\rm HF, S}^{\rm min} &= {U \over 2} +8 {j^2 \over U}+ O\Big({j^4 \over U^3}\Big).
\end{align}
Comparing the expressions in the S and A sectors, we conclude that, in mean field and to leading order in $j/U$, the S-A crossing happens at $j_{13}=2 j^2/|U|$.

\subsubsection{Adiabatic procedure}
Suppose now that we consider the Fermi-Hubbard trimer with $j>0$, $U<0$, $|U| \gg j$ and $\sqrt{3} j^2/|U| < j_{13} <2 j^2/|U|$ and try to identify the ground state with a single spin-up and spin-down particle through adiabatic following. We have just demonstrated that in this situation, the HF solution is in the S sector, while the true ground state is in the A sector. This means that the adiabatic procedure will fail altogether and end up in a symmetric state, which is an excited state of $\mt H$.

\begin{figure}
    \centering
    \includegraphics{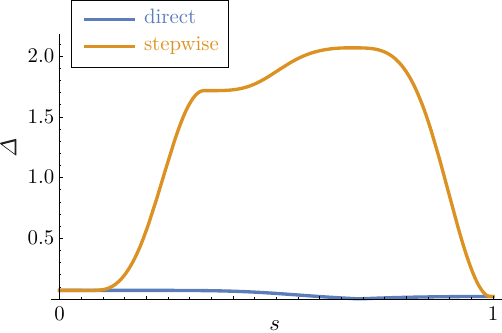}
    \caption{Instantaneous ground state energy gaps in the adiabatic ground state preparation of the Fermi-Hubbard trimer, with $U = -5$, $j = 1$ and $j_{13} = 0.37$. In the direct interpolation, $\H(s) = (1 - s)\H_{\rm HF} + s\H$, a gap closure occurs around $s = 0.69$. The stepwise procedure is carried out as in eq.~\ref{eq:trimer_stepwise_interpolation}, with each step taking a third of the total time. Through symmetry breaking, a gap is visibly opened.}
    \label{fig:trimer_asp}
\end{figure}

Let us now consider how the stepwise procedure works out in this example. The adiabatic procedure starts from the HF hamiltonian, with mean field parameters (called $x$, $y$ in the above) optimised for our choice of $U$, $j$ and $j_{13}$. It is obtained from $\H$ by replacing
\begin{align}
\n_{i\up} \n_{i\down} \rightarrow \n_{i\up} \langle \n_{i\down} \rangle + \langle \n_{i\up} \rangle \n_{i\down} -  \langle \n_{i\up} \rangle \langle \n_{i\down} \rangle.
\end{align}
This implies that the exact two-body hamiltonian $\H^{(2)}$ differs from the two-body HF hamiltonian via diagonal terms only, which directly correspond to the two-body eigenvalues $\lambda_k$ of the stepwise adiabatic following from $\H^{(2)}$ to $\H^{(2)}_{\rm HF}$. Among these eigenvalues, the most negative ones are 
\begin{align}
\lambda_{11} &= \bra{1_\up 1_\down}(\H^{(2)}-\H^{(2)}_{\rm HF})\ket{1_\up 1_\down} = \frac{U}{2}+\frac{32 j^4}{U^3}+O\Big(\frac{j^6}{U^5}\Big)
\nonumber \\[1mm]
\lambda_{33} &= \bra{3_\up 3_\down}(\H^{(2)}-\H^{(2)}_{\rm HF})\ket{3_\up 3_\down} = \lambda_{11}
\end{align}
For $U_1=U_3=U$, these two-body eigenvalues are degenerate, leaving an ambiguity in the definition of the stepwise procedure. After all, one could perfectly define the eigenvectors of the two-particle matrix in such a way that the symmetric and antisymmetric sectors sectors are not mixed. As such, it is necessary to add an arbitrarily small splitting $\delta U = U_3 - U_1$. While this implies that the energy levels in a direct interpolation will not strictly cross, the gap that is opened will only scale in $\delta U$, meaning the time required for the direct interpolation can be made arbitrarily large. Having resolved this ambiguity, we can then design the stepwise adiabatic procedure as follows. Defining
\begin{align}
    \Hr_1 &= \lambda_{11} \atataa{1\up}{1\down}{1\down}{1\up} \nonumber\\
    \Hr_3 &= \lambda_{33} \atataa{3\up}{3\down}{3\down}{3\up} \nonumber\\
    \H^{\rm rest} &= \mt H - \mt H_{\rm HF} - \Hr_1 - \Hr_3
\end{align}
we interpolate thus:
\begin{align}\label{eq:trimer_stepwise_interpolation}
    \Hi &\to \Hi + \Hr_1 \nonumber\\
    &\to \Hi + \Hr_1 + \H^{\rm rest} \nonumber\\
    &\to \Hi + \Hr_1 + \H^{\rm rest} + \Hr_3 = \H.
\end{align}
With this, the discrete symmetry is broken along all steps of the path, and the S-A crossing, which derails the direct adiabatic interpolation from the HF to the exact hamiltonian, is avoided. The resulting development of the instantaneous gap can be seen in figure \ref{fig:trimer_asp}.

\subsection{Fermi-Hubbard model on four sites with alternating hopping}

As a second example, we present a variation on the same theme: a simple model for correlated electrons where a mean field (Hartree Fock) solution is unable to correctly incorporate two-body correlations and as a result puts the ground state in the wrong symmetry sector, derailing adiabatic interpolation with the HF state as starting point. The stepwise adiabatic procedure based on two-body eigenspaces cures this situation.
 
We consider a Fermi-Hubbard model on four sites,
\begin{align*}
\mt H = \includegraphics[width=0.215\textwidth, valign=c]{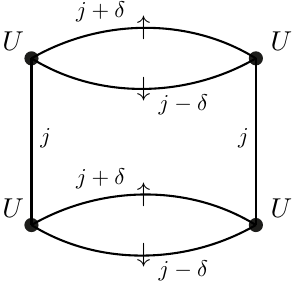}
\end{align*}
 \begin{align}
= \sum_i \sum_\sigma j_{i\sigma} (\a^\dagger_{i\sigma} \a_{i+1,\sigma} + {\rm h.c.}) 
     + U \sum_i \n_{i\up} \n_{i\down},
\end{align}
with a uniform $U<0$ but spin-dependent, non-uniform hoppings
\begin{align}
j_{1\sigma}=j_{3\sigma}=j, ~~ j_{2\up}=j_{4\up}=j+\delta, ~~ j_{2\down}=j_{4\down}=j-\delta.
\end{align} 
Note that we assume periodic boundary conditions, identifying site $i=5$ with $i=1$. We assume half filling, $N_\up=N_\down=2$.

The hamiltonian $\mt H$ is invariant under a reflection $1 \leftrightarrow 4$, $2 \leftrightarrow 3$. Assuming $j>0$ and $0<\delta<j$, it is quickly found that for $U=0$ the spin up  particles have a symmetric (S) ground state, while the ground state for particles with spin down is antisymmetric (A). This renders the overall ground state antisymmetric (A).

\subsubsection{Tracing the exact ground state}

Turning on $U<0$ will change the nature of the ground state, as the particles will tend to form two local pairs. For $|U|\ll j$, there is an effective description in terms of two such pairs with induced pair hopping of order $j^2/U$. It is quickly checked that this leads to a symmetric (S) ground state. The symmetry sector of the many-body ground state will thus change from A to S at a critical value $U_c(j,\delta)<0$.

For a quick estimate of the cross-over point from A to S, we can follow, in (degenerate) first order perturbation theory in $U$ and $\delta$, in the lowest two energy eigenstates (here labeled by their symmetry A or S)
\begin{align}
\langle \mt H \rangle_{\rm A} = -4j -2\delta + U, \qquad  \langle \mt H \rangle_{\rm S} = -4j + {5 \over 4} U,
\end{align}
putting the A-S cross-over at $U_c=-8\delta$.
 
\subsubsection{Mean field (Hartree-Fock, HF) and adiabatic procedure}
 
\begin{figure}
    \centering
    \includegraphics{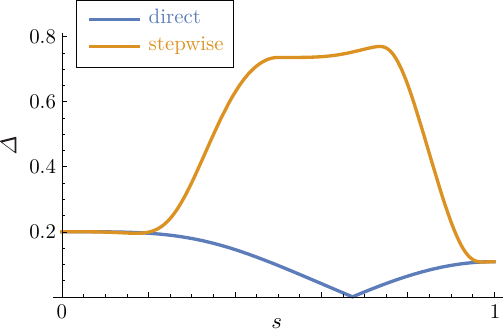}
    \caption{Instantaneous gaps in the direct and stepwise adiabatic ground state preparation of the four-site Fermi-Hubbard model with alternating spins, with $U = -2$, $j = 1$ and $\delta = 0.1$. The direct interpolation causes a level crossing around $s = 0.67$. The stepwise interpolation is carried out by adding a single projector onto one of the sites in the first half, and the rest of $\H - \H_{\rm HF}$ in the second half. It is observed that the stepwise method avoids the level crossing.} 
    \label{fig:four_site_Uneg_asp}
\end{figure}

As in our previous example, a HF state is unable to accommodate the correlations induced by $U<0$. In fact, since all four sites are on equal footing, self-consistent mean fields for both the up and down spins will be uniform on all four sites and will not affect the one-particle states that make up the HF ground state. The mean field ground state will thus be the same as that of the non-interacting problem, and it will be antisymmetric (A). This means that, for $|U|$ larger than $|U_c|$, direct adiabatic interpolation from the Hartree-Fock hamiltonian $\mt{H}_{\rm HF}$ (into which we absorb the uniform mean-field energy shift) to $\mt H$ will result in an excited state of $\mt H$. The stepwise procedure based on two-particle eigenstates of $\mt{H}_{\rm HF} - \mt H$ avoids this problem. It has four non-zero eigenvalues $\lambda_k$, corresponding to projectors on each of the sites. (As in the three-site example, for this to be unambiguous one needs to assume an arbitrarily small splitting of the value of $U$ for the four sites.) Adding one of these in the first step and the other three in the second step gives an adiabatic path that breaks the left-right symmetry and thereby avoids the crossing, allowing a correct interpolation from the antisymmetric HF state to the symmetric ground state of $\mt H$. This can be seen in figure \ref{fig:four_site_Uneg_asp}.
\begin{figure}
    \centering
    \includegraphics{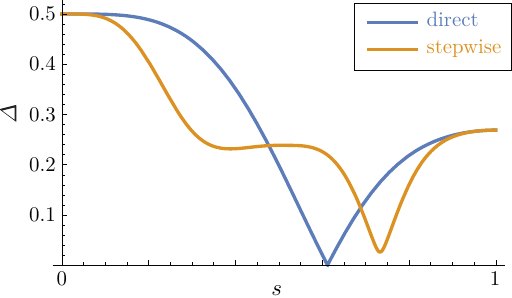}
    \caption{Instantaneous gaps for the four-site Fermi-Hubbard model, with positive $U$ ($U=+2$, $j=1$ and $\delta=0.25$). In the same way as in figure \ref{fig:four_site_Uneg_asp}, the direct interpolation causes a gap closure around $s = 0.61$ while the stepwise procedure keeps the gap open. Note however a small gap in the stepwise procedure around $s = 0.67$.}
    \label{fig:four_site_Upos_asp}
\end{figure}
We observe that in the same model with $U>0$ the stepwise adiabatic following similarly avoids the problem of an exact A-S crossing, but in that case the gaps coming with the stepwise procedure are smaller and tend to decrease with $\delta$. This is seen in figure \ref{fig:four_site_Upos_asp}.

\section{Complexity considerations\label{sec:opnorm}}
Having seen the potential of a (partially) piecewise interpolation to lift gap closures, we will now study more generally the worst-case complexity of direct and piecewise paths, in view of the adiabatic complexity bound of eq.~\ref{eq:JRS_T}. For simplicity, we replace the gap $\Delta(s)$ by its minimum $\Delta = \min_{s\in[0, 1]}\Delta(s)$, and consider complexity in terms of this parameter. What remains then is to determine the scaling of the numerators
\begin{align}\label{eq:adiabatic_numerator}
    I_n := \int_0^1 \|\partial^n_s{\mt H}(\sigma)\|^{2/n} d\sigma \quad (n \in \{1, 2\})
\end{align}
in the system size. Since for typical systems of interest, the number of particles $N$ scales proportionally with the number of one-particle modes $L$, we take $L$ as the system size scaling parameter.\par
Now, from eqs.~\ref{eq:newpath} and \ref{eq:nqf_1st_decomposition},
\begin{align}
    \|\partial^n_s\H(s)\| \leq \sum_{k=1}^{\L(\L-1)/2} |\partial^n_s\gamma_k(s)| \cdot |\lambda_k| \cdot \|\mtgPhi_k\|.
\end{align}
Note that the path functions $\gamma_k$ can always be chosen such that $|\partial^n_s \gamma_k(s)|$ is upper bounded by a constant. In particular, one may always pick all $\gamma_k$ such that $\partial_s \gamma_k(0) = \partial_s \gamma_k(1) = 0$, so that the boundary term in eq.~\ref{eq:JRS_T} drops out. The two-body eigenvalues $\lambda_k$ then, being the eigenvalues of $\mt F$, are bounded by the energy scale of a two-particle system which does not grow with the system size. Therefore the norms $\|\mtgPhi_k\|$ are the only meaningful quantities to be upper bounded. As such, it suffices to consider only the first derivative numerator $I_1$. We shall universally upper bound the operator norm of any pseudoprojector $\mtgPhi_k$ in the following, and shall henceforth drop the subscript $k$. Afterwards, we discuss some implications of this bound for different choices of paths and systems.
\subsubsection{Upper bound to the pseudoprojector operator norm}
Define $\mt b$ such that $\mtgPhi = \mt b\t\mt b$ for some operator $\mtgPhi$ from the decomposition. When we define the $\L\times\L$ antisymmetric matrix $\tilde{\mtg\phi}$ with entries
\begin{align}
    \tilde\phi^{PR} := \phi^{(PR)} - \phi^{(RP)}
\end{align}
we may write
\begin{align}
    \mt b\t = \sum_{P<R} \phi^{(PR)} \a_P\t\a_R\t = \frac12\sum_{PR} \tilde \phi^{PR} \a_P\t\a_R\t.
\end{align}
Next, apply a Youla decomposition $\tilde{\mtg\phi} = \mt V\mtgXi\mt V\T$ where $\mt V$ is an $\L \times \L$ orthogonal matrix and
\begin{align}\label{eq:phi_k_asymm_decomp}
    \mtgXi = \bigoplus_{m=1}^{\L/2} \mat{0 & \xi_m \\ -\xi_m & 0}
\end{align}
if $\L$ is even; if $\L$ is odd, $\mtgXi$ has an additional row and column of zeros. This then yields
\begin{align}
    \mt b\t = \sum_{m=1}^{\lfloor\L/2\rfloor} \xi_m \tilde\a_{2m-1}\t\tilde\a_{2m}\t \label{eq:W_normal_form}
\end{align}
where we defined the rotated fermionic operators $\tilde\a_K^{(\dagger)} = \sum_P V_{PK} \a_P^{(\dagger)}$. Note that since the vector with entries $\phi^{(PR)}$ is a normalised eigenvector, the squares of $\xi_m$ sum to unity.\par
Since $\|\mtgPhi\| = \|\mt b\|^2 = \max_{\ket\Psi}\|\mt b\ket\Psi\|^2$, in order to obtain the spectral norm of $\mtgPhi$ it is sufficient to find the state whose norm is maximised under the application of $\mt b$. Now, from eq.~\ref{eq:W_normal_form} we observe that $\mt b$ defines a set of pairs $\{(2m - 1, 2m)\}_{m=1}^{\floor{\L/2}}$, and only annihilates particles from a product state $\prod_P\tilde{\a}\t_P\ket{\rm vac}$ if they appear together in these pairs. As such, it makes sense to describe a product state in terms of its fermion pairs and its unpaired fermions. We shall denote a product state as a ket $\ket{\P, \U}$ where $\P$ is the set of filled pairs and $\U$ is the set of remaining unpaired fermions; in other words,
\begin{align}\label{eq:ketPUdef}
    \ket{\P, \U} = \bigg( \kern0.1333em {\prod_{i\in\U}}' \tilde{\a}_i\t \bigg) \bigg( \prod_{m\in\P} \tilde\a_{2m-1}\t\tilde\a_{2m}\t \bigg) \ket{\rm vac}
\end{align}
where the prime on the leftmost product symbol indicates that a certain order of the unpaired modes is assumed, in order to fix the sign of $\ket{\P, \U}$. In this notation, the matrix elements of $\mtgPhi$ are given by
\begin{align}\label{eq:Phi_matrix_element}
    \bra{\P, \U} \mtgPhi \ket{\P', \U'} = \delta_{\U\U'} \begin{cases}
        \sum_{m \in \P} \xi_m^2 \hspace*{-0.2cm} & \text{if } \P = \P'\\
        \xi_m\xi_n & \text{if } \P\setminus\!\{m\} = \P'\setminus\!\{n\}\\
        0 & \text{otherwise.}
    \end{cases}
\end{align}
From eq.~\ref{eq:Phi_matrix_element}, it is clear that $\|\mt b\ket{\P, \U}\|$ is maximised when $\ket{\P, \U}$ lies in the sector with a minimal number of unpaired fermions (zero if $N$ is even, one if odd). Furthermore, $\b\t\b$ preserves the unpaired fermions, and therefore the state $\ket\Psi$ that maximises the norm must lie in this sector.\par
If $N$ is even, all particles in this sector are paired up. Such paired fermions, then, are equivalent to what are known as \emph{hardcore bosons} (HCBs): particles whose operator algebra commutes at different sites, but which may only singly occupy any given site. In this sense, the set $\{\b_m\t\ket{\rm vac}\}_{m=1}^{\lfloor\L/2\rfloor}$ with $\b_m\t = \tilde\a_{2m-1}\t\tilde\a_{2m}\t$ may be viewed as a single-particle HCB basis, and $\mt b\t$, to which we shall add a subscript $\mt b\t = \b_{\vcg\xi}\t$, is then a rotated HCB creation operator. A universal upper bound on the spectral norm of a pseudoprojector $\mtgPhi$ is now given by
\begin{align}
    \|\mtgPhi\| \leq \max_{\|\vcg\xi\|=1} \max_{\ket\Psi\in\mathcal H^{\rm HCB}_{\lfloor\L/2\rfloor, N/2}} \bra\Psi\b_{\vcg\xi}\t\b_{\vcg\xi}\ket\Psi \label{eq:specop_Q_upper_bound}
\end{align}
where $\mathcal H^{\rm HCB}_{l, n}$ denotes an $l$-site, $n$-particle HCB Hilbert space. An expression for the right-hand side of eq.~\ref{eq:specop_Q_upper_bound} was found by Tennie et al. \cite[theorem 1]{Tennie2017}; the upper bound that follows is
\begin{theorem}\label{prop:Q_upper_bound_N_even}
For even particle number $N$, a universal upper bound on the operator norm of a pseudoprojector $\mtgPhi$ is given by
\begin{align}
    \|\mtgPhi\| \leq \frac{N/2}{\floor{\L/2}}(\floor{\L/2} - N/2 + 1). \label{eq:upper_bound_Q_even}
\end{align}
\end{theorem}
For $N$ odd, a similar result may be found through the observation that any eigenstate of $\mtgPhi$ with maximum eigenvalue must lie in the sector where all fermions except one are paired up, for the same reason as discussed above. The problem of upper bounding the operator norm of $\|\mtgPhi\|$ is then equivalent to the even case in a Hilbert space with one less HCB site available. This is formalised in the following theorem.
\begin{theorem}\label{prop:Q_upper_bound_N_odd}
For odd $N$, $\|\mtgPhi\|$ is upper bounded by
\begin{align}
    \|\mtgPhi\| \leq \frac{\floor{N/2}}{\floor{\L/2} - 1}(\floor{\L/2} - \floor{N/2}). \label{eq:upper_bound_Q_odd}
\end{align}
\end{theorem}
The bounds of theorems \ref{prop:Q_upper_bound_N_even} and \ref{prop:Q_upper_bound_N_odd} are also tight.
\begin{theorem}\label{prop:even_odd_bounds_tight}
    Let $\ell = \floor{\L/2}$. The upper bound in the even case, theorem \ref{prop:Q_upper_bound_N_even}, is saturated by taking $\xi_m = 1/\sqrt\ell \; \forall m$, and taking $\ket\Psi$ to be the maximally symmetric state
    \begin{align}
        \ket\Psi = \binom{\ell}{N/2}^{-1/2} \sum_{\P:|\P|=N/2} \ket{\P, \emptyset}
    \end{align}
    where the sum runs over all sets $\P$ of $N/2$ HCB sites.\par
    The upper bound of in the odd case, theorem \ref{prop:Q_upper_bound_N_odd}, is saturated by $\xi_m = 1/\sqrt{\ell - 1} \; \forall m\neq \ell$, $\xi_\ell = 0$ and
    \begin{align}
        \ket\Psi = \binom{\ell - 1}{\floor{N/2}}^{-1/2} \sum_{\substack{\P:|\P|=\floor{N/2} \\ \ell\notin \P}} \ket{\P, \{2\ell - 1\}}.
    \end{align}
\end{theorem}
The proofs of theorems \ref{prop:Q_upper_bound_N_odd} and \ref{prop:even_odd_bounds_tight} are deferred to appendices \ref{app:proof_prop_Q_upper_bound_N_odd} and \ref{app:proof_prop_even_odd_bounds_tight} respectively. We note that instead of the $\ell$-th HCB site, the unpaired fermion could occupy any HCB site.\par
In typical systems of interest, the particle number $N$ will scale proportionally to the number of modes $\L$; we have thus shown that the operator norm of each pseudoprojector $\mtgPhi_k$ scales at most linearly in $\L$.
\subsubsection{Implications}
Let us now think about how the above result can be used to reason about the adiabatic complexity of a choice of system or path, in terms of the numerator of eq.~\ref{eq:adiabatic_numerator}. We set a baseline with the following bound which applies to direct interpolation. Define the residual hamiltonian $\Hr$ in a generic fashion as in eq.~\ref{eq:generic_H_antisymmetrised}, and observe that
\begin{align}
    \|\Hr\| &\leq \sum_{\substack{P<R\\Q<S}} \|F_{(PR)(QS)} \, \atataa PRSQ\| \nonumber\\
    &\leq \sum_{\substack{P<R\\Q<S}} |F_{(PR)(QS)}| \leq \sqrt{\L(\L - 1)/2} \; \|\mt F\|_{\rm F} \label{eq:naive_upper_bound_Hr}
\end{align}
where $\|\cdot\|_{\rm F}$ denotes the Frobenius norm. Given the dimensionality of $\mt F$, it is clear that $\|\mt F\|_{\rm F} \leq c\sqrt{\L(\L - 1)/2}$ for some nonnegative constant $c$, and thus $\|\Hr\| \leq O(\L^2)$. The resulting numerator, for direct interpolation, from eq.~\ref{eq:adiabatic_numerator} then scales as $O(\L^4)$.\par
In comparison, consider a fully stepwise scheme where we ``adiabatically add'' every term $\Hr_k$ from the eigendecomposition, eq.~\ref{eq:nqf_1st_decomposition}, separately, i.e. we evolve
\begin{align}
    \Hi \to \Hi + \Hr_{k_1} \to \Hi + \Hr_{k_1} + \Hr_{k_2} \to \cdots \to \H \label{eq:stepwise_adiabatic_addition}
\end{align}
where $\mt A \to \mt B$ denotes a direct adiabatic interpolation between $\mt A$ and $\mt B$. In this scheme, at any point in the evolution, exactly one of the $\gamma_k$ (cf. eq.~\ref{eq:newpath}) must increase at a rate scaling in the number of terms (which is $O(\L^2)$), with the rest staying constant (being either 0 or 1). From the universal result that $\|\mtgPhi_k\| \leq O(\L)$, we then have
\begin{align}
    I_1 \leq \int_0^1 |O(\L^2)O(\L)|^2 d\sigma = O(\L^6). \label{eq:stepwise_O(L^6)}
\end{align}
This indicates that the fully stepwise procedure is unfavourable as compared to direct interpolation. This is no surprise: with the work of Tomka et al \cite{Tomka2016} in mind, the direct path is a geodesic if the gap is held constant, whereas the fully stepwise approach is a walk along the corners of a hypercube in parameter space.\par
However, we emphasise that these bounds are worst-case and can be improved in certain settings. Consider, for example, the standard Fermi-Hubbard model of eq.~\ref{eq:fermi-hubbard} with the hopping part plus the chemical potential (which is proportional to the identity in for fixed $N$) taken as the initial hamiltonian. The spectral norm of the residual hamiltonian $\Hr$ is easily found to be $UN/2 = O(\L)$, leading to $I_1 \leq O(\L^2)$. Furthermore, since $\Hr$ is diagonal in the two-particle position basis, all its pseudoprojectors are of the form $\mtgPhi_k = \atataa{I_k}{J_k}{J_k}{I_k}$ and thus have unit spectral norm. Since $\Hr$ contains $\L/2$ such terms, a fully stepwise procedure yields the same numerator bound, $I_1 \leq O(\L^2)$.\par
On the other hand, the paired fermion formalism may be used to find examples which saturate the bounds of both eq.~\ref{eq:naive_upper_bound_Hr} and \ref{eq:stepwise_O(L^6)}. To this end, we define a residual hamiltonian $\Hr$ and the corresponding operators $\mtgPhi_k$ through its two-body eigenstates. As we have seen, a fully paired state $\ket\Psi = (\xi_1 \a\t_1\a\t_2 + \cdots + \xi_{\floor{\L/2}}\a\t_{\L - 1}\a\t_{\L})\ket{\rm vac}$, where $\xi_m = (\floor{\L/2})^{-1/2}$ for all $m$, gives rise to a pseudoprojector with a maximal norm that is $O(\L)$. Since the one-body basis is free to choose, we can permute the one-body modes in $\L - 1$ ways such that all resulting two-body states are fully paired and mutually orthogonal\footnote{One way to see this is to draw a complete graph of $\L$ nodes where nodes $2, \ldots, \L$ are drawn in a circle around node 1. A full pairing (also known as perfect matching) may then be found by selecting an edge from node 1 to any other node and pairwise connecting the other nodes through edges orthogonal to the first edge. In this way, we find $\L - 1$ pairings without drawing any parallel edges, guaranteeing that that resulting two-body states are mutually orthogonal.}. Furthermore, we make use of the fact that the mapping $\xi_m \mapsto -\xi_m$ preserves the spectrum of any pseudoprojector $\mtgPhi$. After all, from eq.~\ref{eq:Phi_matrix_element} we see that $\xi_m$ appears in an off-diagonal element $\bra{\P, \U}\mtgPhi\ket{\P', \U}$ only if $m \in \P$ and $m \notin \P'$ or vice versa. Therefore this mapping is realised by the transformation $\mtgPhi_k \mapsto \mtgSigma \mtgPhi_k \mtg\Sigma$ where $\mtgSigma$ is a diagonal matrix with a $-1$ entry in those columns corresponding to the product state $\ket{\P, \U}$ where $m \in \P$, and a $+1$ entry elsewhere. Now we can use this to vary the signs of the terms in a fully paired two-body state; if $\L$ is a power of two, we can construct $\L/2$ vectors $\vcg\xi$ that are the (normalised) columns of an Hadamard matrix, so that the resulting $\L/2$ two-body states are mutually orthogonal. As such, we have defined the $\L(\L - 1)/2$ two-body eigenstates necessary to describe an interacting hamiltonian, each of which fully paired.\par
Now, since we have $\L(\L - 1)/2$ pseudoprojectors with maximal norm, the inequality of eq.~\ref{eq:stepwise_O(L^6)} is automatically saturated (for any choice of the two-body eigenvalues), and the adiabatic numerator is maximised for the stepwise procedure. Furthermore, this construction also attains a maximally scaling numerator in the case of direct interpolation, if we set all two-body eigenvalues to 1. Indeed, consider the sum over all pseudoprojectors $\mtgPhi_k$ which carry the same pairing (and therefore only differ in their $\vcg\xi$ vectors):
\begin{align}
    \sum_{k\,:\,\text{same pairing}} \mtgPhi_k &= \sum_k \sum_{mn} \xi^k_m\xi^k_n \, \atataa{2m-1}{2m}{2n}{2n-1} \nonumber\\[1mm]
    & = \sum_m \n_{2m-1}\n_{2m}.
\end{align}
In other words, this particular sum of pseudoprojectors is an operator that counts all pairs of fermions in a product state that coincide with the mode pairs that define the pseudoprojectors. As a result, the sum over \emph{all} pseudoprojectors defined in this example is an operator that counts all possible pairs of fermions, and is therefore simply equal to $N(N - 1)/2$ times the identity. This operator saturates the bound $\|\Hr\| \leq O(\L^2)$ (under the assumption that $\L/N = O(1)$) and therefore realises a maximal adiabatic numerator scaling of $O(\L^4)$. This analysis establishes a condition, expressed in the paired fermion formalism, on the two-body eigenstates that yields worst-case numerator scaling for both direct and fully stepwise interpolation. In addition, it shows that direct interpolation indeed outperforms a fully stepwise protocol in this sense.

\section{Conclusion}\label{sec:discussion}
In this work, we have proposed a new protocol for adiabatic preparation of fermionic many-body ground states based on the eigendecomposition of the (combined one- and two-body) coefficient tensor of the residual hamiltonian, being the difference between the initial and final hamiltonian, in second quantisation. The eigenvectors in this decomposition are equivalent to two-body eigenstates of the residual hamiltonian. The method decomposes the residual hamiltonian into a sum of simpler terms, each of which corresponds to an eigenvalue and eigenvector from the eigendecomposition. In the adiabatic scheme, every point along the evolution path is then a linear combination of these terms.\par
We have demonstrated how this idea may be applied to generalised Fermi-Hubbard models, through a few small worked examples. Our finding is that a level crossing occurring in a direct interpolation from a mean-field hamiltonian, which arises from a discrete one-body symmetry, can be cured with the two-body decomposition approach. Although this is not a general superiority result, it shows the existence of scenarios in which the use of (partially) piecewise paths resulting from a two-body decomposition is advantageous as compared to direct interpolation. More precisely, in this approach, one can design a procedure which explicitly breaks the symmetry by interpolating through an intermediate hamiltonian which contains only a subset of the hamiltonian terms from the decomposition. As a result, a gap is seen to be opened. The conditions for this to occur are rather specific: while the initial hamiltonian must share a symmetry with the target hamiltonian and place the ground state in the incorrect symmetry sector, the two-particle matrix of the residual hamiltonian must have degenerate eigenvalues in order to mix two-particle eigenstates from the relevant sectors.\par
Having established this gap opening potential of the two-body decomposition methodt, we proceeded to analyse the adiabatic complexity of piecewise paths more broadly, by examining how the two-body decomposition influences the numerator part of the complexity of many-body adiabatic state preparation. This numerator is primarily dependent on the operator norm of each term from the residual hamiltonian decomposition. We have found that a description in terms of fermion pairs (or equivalently, hard-core bosons) is key to understanding the scaling, in terms of the number of single-particle modes $L$, of this operator norm and therefore the adiabatic numerator. The main result is that each residual hamiltonian term scales at most as $O(L)$, for a typical system where the number of particles $N$ scales proportionally with $L$. This result has different implications, depending on the system under investigation and the chosen evolution path. For example, for the Fermi-Hubbard model with the interaction part taken as the residual hamiltonian, the adiabatic complexity scales as $O(L^2/\Delta^3)$ both in a direct interpolation, and when following a fully piecewise path. This is due to the fact that the norm of each residual hamiltonian term from the decomposition scales as $O(1)$, and there are only $L$ nonzero terms. On the other hand, for a situation in which all two-body eigenstates are uniformly weighted superpositions of distinct fermion pairs, each term attains the maximal scaling of $O(L)$; as a result, the time complexity of direct interpolation in this case scales as $O(L^4/\Delta^3)$, whereas under a fully stepwise path the we find an $O(L^6/\Delta^3)$ scaling. Both these scalings are worst-case. This finding agrees with the statement by Tomka et al. that a geodesic path in parameter space is generally beneficial in terms of time complexity \cite{Tomka2016}. The result suggests that one should be selective when choosing which of the hamiltonian terms from the decomposition to interpolate in a piecewise fashion, and which to interpolate directly. Namely, by piecewise interpolating only those terms which (are expected to) break any relevant symmetries, one retains the power to lift level crossings, while avoiding potentially unfavourable scaling in $L$. We note however that the situation of maximal scaling is a case where the residual hamiltonian is particularly dense. In large chemical systems, for example, the two-electron part of the hamiltonian is typically sparse, so it is expected that a lower adiabatic numerator can be achieved for such systems.\par
All in all, our examples show that the two-body eigendecomposition method can outperform direct interpolation through symmetry breaking, and we expect that the method can be helpful in situations beyond a single reflection symmetry. An example is the nonrelativistic treatment of molecular electronic structure, which maintains a $SU(2)$ spin symmetry. Another approach could be the use of a two-body eigendecomposition as a black box if there is a hidden symmetry and the precise cause of a gap closure is not straightforward to determine.

\section{Acknowledgments}
We thank Luuk Visscher, Emiel Koridon and Stefano Polla for valuable discussions.\par
This work was supported by the Dutch Ministry of Economic Affairs and Climate Policy (EZK), as part of the Quantum Delta NL programme.

\bibliographystyle{unsrt}
\bibliography{biblio}

\begin{thebibliography}{10}

\bibitem{Feynman1982}
Richard~P. Feynman.
\newblock Simulating physics with computers.
\newblock {\em International Journal of Theoretical Physics}, 21:467--488, 6
  1982.

\bibitem{Lloyd1996}
Seth Lloyd.
\newblock Universal quantum simulators.
\newblock {\em Science}, 273:1073--1078, 8 1996.

\bibitem{Kitaev1995}
Alexei~Y. Kitaev.
\newblock Quantum measurements and the abelian stabilizer problem.
\newblock {\em Electron. Colloquium Comput. Complex.}, TR96, 11 1995.

\bibitem{Whitfield2011}
James~D. Whitfield, Jacob Biamonte, and Alan Aspuru-Guzik.
\newblock Simulation of electronic structure hamiltonians using quantum
  computers.
\newblock {\em Molecular Physics}, 109:735--750, 3 2011.

\bibitem{Low2016}
Guang~Hao Low and Isaac~L. Chuang.
\newblock Hamiltonian simulation by qubitization.
\newblock {\em Quantum}, 3, 10 2016.

\bibitem{Babbush2018a}
Ryan Babbush, Nathan Wiebe, Jarrod McClean, James McClain, Hartmut Neven, and
  Garnet Kin~Lic Chan.
\newblock Low-depth quantum simulation of materials.
\newblock {\em Physical Review X}, 8:011044, 3 2018.

\bibitem{Babbush2018b}
Ryan Babbush, Craig Gidney, Dominic~W. Berry, Nathan Wiebe, Jarrod McClean,
  Alexandru Paler, Austin Fowler, and Hartmut Neven.
\newblock Encoding electronic spectra in quantum circuits with linear {T}
  complexity.
\newblock {\em Physical Review X}, 8:041015, 10 2018.

\bibitem{Kivlichan2020}
Ian~D. Kivlichan, Craig Gidney, Dominic~W. Berry, Nathan Wiebe, Jarrod McClean,
  Wei Sun, Zhang Jiang, Nicholas Rubin, Austin Fowler, Alán Aspuru-Guzik,
  Hartmut Neven, and Ryan Babbush.
\newblock Improved fault-tolerant quantum simulation of condensed-phase
  correlated electrons via trotterization.
\newblock {\em Quantum}, 4:296, 7 2020.

\bibitem{Lee2021}
Joonho Lee, Dominic~W. Berry, Craig Gidney, William~J. Huggins, Jarrod~R.
  McClean, Nathan Wiebe, and Ryan Babbush.
\newblock Even more efficient quantum computations of chemistry through tensor
  hypercontraction.
\newblock {\em PRX Quantum}, 2:030305, 9 2021.

\bibitem{VonBurg2021}
Vera von Burg, Guang~Hao Low, Thomas Häner, Damian~S. Steiger, Markus Reiher,
  Martin Roetteler, and Matthias Troyer.
\newblock Quantum computing enhanced computational catalysis.
\newblock {\em Physical Review Research}, 3:033055, 9 2021.

\bibitem{Cade2019}
Chris Cade, Lana Mineh, Ashley Montanaro, and Stasja Stanisic.
\newblock Strategies for solving the {Fermi-Hubbard} model on near-term quantum
  computers.
\newblock {\em Physical Review B}, 102, 12 2019.

\bibitem{Montanaro2020}
Ashley Montanaro and Stasja Stanisic.
\newblock Compressed variational quantum eigensolver for the {Fermi-Hubbard}
  model.
\newblock {\em arXiv: Quantum Physics}, 6 2020.

\bibitem{Wei2020}
Shijie Wei, Hang Li, and GuiLu Long.
\newblock A full quantum eigensolver for quantum chemistry simulations.
\newblock {\em Research}, 2020, 1 2020.

\bibitem{Tilly2022}
Jules Tilly, Hongxiang Chen, Shuxiang Cao, Dario Picozzi, Kanav Setia, Ying Li,
  Edward Grant, Leonard Wossnig, Ivan Rungger, George~H. Booth, and Jonathan
  Tennyson.
\newblock The variational quantum eigensolver: A review of methods and best
  practices.
\newblock {\em Physics Reports}, 986:1--128, 11 2022.

\bibitem{Farhi2000}
Edward Farhi, Jeffrey Goldstone, Sam Gutmann, and Michael Sipser.
\newblock Quantum computation by adiabatic evolution.
\newblock {\em arXiv: Quantum Physics}, 1 2000.

\bibitem{Wecker2015}
Dave Wecker, Matthew~B. Hastings, Nathan Wiebe, Bryan~K. Clark, Chetan Nayak,
  and Matthias Troyer.
\newblock Solving strongly correlated electron models on a quantum computer.
\newblock {\em Physical Review A - Atomic, Molecular, and Optical Physics}, 92,
  6 2015.

\bibitem{Perez2022}
Axel Pérez-Obiol, Adrián Pérez-Salinas, Sergio Sánchez-Ramírez, Bruna~G.M.
  Araújo, and Artur Garcia-Saez.
\newblock Adiabatic quantum algorithm for artificial graphene.
\newblock {\em Physical Review A}, 106:052408, 11 2022.

\bibitem{Du2010}
Jiangfeng Du, Nanyang Xu, Xinhua Peng, Pengfei Wang, Sanfeng Wu, and Dawei Lu.
\newblock {NMR} implementation of a molecular hydrogen quantum simulation with
  adiabatic state preparation.
\newblock {\em Physical Review Letters}, 104:030502, 1 2010.

\bibitem{Veis2014}
Libor Veis and Jiří Pittner.
\newblock Adiabatic state preparation study of methylene.
\newblock {\em The Journal of Chemical Physics}, 140, 6 2014.

\bibitem{Babbush2014}
Ryan Babbush, Peter~J. Love, and Alan Aspuru-Guzik.
\newblock Adiabatic quantum simulation of quantum chemistry.
\newblock {\em Scientific Reports 2014 4:1}, 4:1--11, 10 2014.

\bibitem{Sugisaki2022}
Kenji Sugisaki, Kazuo Toyota, Kazunobu Sato, Daisuke Shiomi, and Takeji Takui.
\newblock Adiabatic state preparation of correlated wave functions with
  nonlinear scheduling functions and broken-symmetry wave functions.
\newblock {\em Communications Chemistry}, 5:1--13, 7 2022.

\bibitem{Jansen2006}
Sabine Jansen, Mary-Beth Ruskai, and Ruedi Seiler.
\newblock Bounds for the adiabatic approximation with applications to quantum
  computation.
\newblock {\em Journal of Mathematical Physics}, 48:102111--102111, 3 2006.

\bibitem{Suzuki1993}
Masuo Suzuki.
\newblock General decomposition theory of ordered exponentials.
\newblock {\em Proceedings of the Japan Academy, Series B}, 69:161--166, 1993.

\bibitem{Childs2019}
Andrew~M. Childs, Yuan Su, Minh~C. Tran, Nathan Wiebe, and Shuchen Zhu.
\newblock A theory of trotter error.
\newblock {\em Physical Review X}, 11, 12 2019.

\bibitem{Wan2022}
Kianna Wan and Isaac~H Kim.
\newblock Fast digital methods for adiabatic state preparation.
\newblock {\em arXiv: Quantum Physics}, 4 2022.

\bibitem{Low2019}
Guang~Hao Low and Nathan Wiebe.
\newblock Hamiltonian simulation in the interaction picture.
\newblock {\em arXiv: Quantum Physics}, 5 2019.

\bibitem{Kieferova2018}
Maria Kieferova, Artur Scherer, and Dominic Berry.
\newblock Simulating the dynamics of time-dependent hamiltonians with a
  truncated {Dyson} series.
\newblock {\em Physical Review A}, 99, 5 2018.

\bibitem{Aharonov2003}
Dorit Aharonov and Amnon Ta-Shma.
\newblock Adiabatic quantum state generation and statistical zero knowledge.
\newblock {\em Conference Proceedings of the Annual ACM Symposium on Theory of
  Computing}, pages 20--29, 1 2003.

\bibitem{Lemieux2021}
Jessica Lemieux, Artur Scherer, and Pooya Ronagh.
\newblock Reflection-based adiabatic state preparation.
\newblock {\em arXiv: Quantum Physics}, 11 2021.

\bibitem{Boixo2009}
Sergio Boixo, Emanuel Knill, and Rolando Somma.
\newblock Eigenpath traversal by phase randomization.
\newblock {\em Quantum Inf. Comput.}, 9:833--855, 2009.

\bibitem{Boixo2010}
S.~Boixo, E.~Knill, and R.~D. Somma.
\newblock Fast quantum algorithms for traversing paths of eigenstates.
\newblock {\em arXiv: Quantum Physics}, 5 2010.

\bibitem{Albash2016}
Tameem Albash and Daniel~A. Lidar.
\newblock Adiabatic quantum computing.
\newblock {\em Reviews of Modern Physics}, 90, 11 2016.

\bibitem{Tomka2016}
Michael Tomka, Tiago Souza, Steven Rosenberg, and Anatoli Polkovnikov.
\newblock Geodesic paths for quantum many-body systems.
\newblock {\em arXiv: Quantum Gases}, 6 2016.

\bibitem{Demirplak2003}
Mustafa Demirplak and Stuart~A. Rice.
\newblock Adiabatic population transfer with control fields.
\newblock {\em Journal of Physical Chemistry A}, 107:9937--9945, 11 2003.

\bibitem{Demirplak2005}
Mustafa Demirplak and Stuart~A. Rice.
\newblock Assisted adiabatic passage revisited†.
\newblock {\em Journal of Physical Chemistry B}, 109:6838--6844, 4 2005.

\bibitem{Demirplak2008}
Mustafa Demirplak and Stuart~A. Rice.
\newblock On the consistency, extremal, and global properties of
  counterdiabatic fields.
\newblock {\em The Journal of Chemical Physics}, 129:154111, 10 2008.

\bibitem{Rubin2022}
Nicholas~C. Rubin, Joonho Lee, and Ryan Babbush.
\newblock Compressing many-body fermion operators under unitary constraints.
\newblock {\em Journal of Chemical Theory and Computation}, 18:1480--1488, 3
  2022.

\bibitem{Koch2003}
Henrik Koch, Alfredo Sánchez~De Merás, and Thomas~Bondo Pedersen.
\newblock Reduced scaling in electronic structure calculations using {Cholesky}
  decompositions.
\newblock {\em The Journal of Chemical Physics}, 118:9481, 5 2003.

\bibitem{Nottoli2021}
Tommaso Nottoli, Jürgen Gauss, and Filippo Lipparini.
\newblock Second-order {CASSCF} algorithm with the {Cholesky} decomposition of
  the two-electron integrals.
\newblock {\em Journal of Chemical Theory and Computation}, 17:6819--6831, 11
  2021.

\bibitem{Roeggen2008}
I.~Røeggen and Tor Johansen.
\newblock {Cholesky} decomposition of the two-electron integral matrix in
  electronic structure calculations.
\newblock {\em The Journal of Chemical Physics}, 128:194107, 5 2008.

\bibitem{Francis2022}
Akhil Francis, Ephrata Zelleke, Ziyue Zhang, Alexander~F. Kemper, and James~K.
  Freericks.
\newblock Determining ground-state phase diagrams on quantum computers via a
  generalized application of adiabatic state preparation.
\newblock {\em Symmetry}, 14:809, 4 2022.

\bibitem{Tennie2017}
Felix Tennie, Vlatko Vedral, and Christian Schilling.
\newblock Universal upper bounds on the bose-einstein condensate and the
  hubbard star.
\newblock {\em Physical Review B}, 96, 7 2017.

\end{thebibliography}

\clearpage

\appendix

\section{Proof of theorem \ref{prop:Q_upper_bound_N_odd}}\label{app:proof_prop_Q_upper_bound_N_odd}
The proof largely follows that of Tennie et al. \cite[theorem 1]{Tennie2017}.\par
We seek to maximise the expectation value $\bra\Psi\b_{\vcg\xi}\t \b_{\vcg\xi}\ket\Psi$ with respect to $\ket\Psi$ and $\vcg\xi$. The hard-core boson (HCB) operator $\b_{\vcg\xi}$ is defined as $\b_{\vcg\xi} = \sum_m \xi_m \b_m = \sum_m \xi_m \tilde\a_{2m}\tilde\a_{2m-1}$; in the following, we will use $m$ both as an index of HCB sites and as a shorthand for the (equivalent) pair of fermionic sites $(2m - 1, 2m)$. Additionally, we use the \emph{flattening} symbol $\phi$ to convert between sets of pairs and sets of fermionic sites,
\begin{align}
    \phi(\{(\mu_1, \mu_2), \ldots\}) = \{\mu_1, \mu_2, \ldots\}.
\end{align}
In the subspace of maximally paired $N$-particle states (with a single fermion left unpaired), $\ket\Psi$ may be expanded as
\begin{align}
    \ket\Psi = \sum_{I, i} A_{I, i}\ket{I, \{i\}}
\end{align}
where $I$ is a set of fermionic pairs and $\ket{I, \{i\}} = \tilde\a_i\t \prod_{m\in I} \b_m\t \ket{\rm vac}$, in line with eq.~\ref{eq:ketPUdef}. Furthermore, we \emph{define} $A_{I, i} = 0$ if $i \in \phi(I)$ or $|I| \neq \floor{N/2}$.
The desired expectation value may then be expressed as
\begin{align}\label{eq:ex_val_inner_prod}
    \bra\Psi\b_{\vcg\xi}\t\b_{\vcg\xi}\ket\Psi &= \sum_{\substack{I, i\\J, j}} \sum_{mn} A_{I, i} A_{J, j} \, \xi_m \xi_n \, \bra{I, \{i\}} \b_m\t\b_n \ket{J, \{j\}} \nonumber\\
    &= \sum_{\substack{I, i\\J, j}} A_{I, i} A_{J, j} \sum_{\substack{m\in I\\n\in J}} \xi_m \xi_n \, \delta_{I\setminus\{m\}, J\setminus\{n\}} \, \delta_{ij} \nonumber\\
    &= \sum_{I', i} \sum_{mn} A_{I'\cup\{m\}, i} \, A_{I'\cup\{n\}, i} \, \xi_m \xi_n \nonumber\\
    &= \sum_{I', i} \Big( \sum_m A_{I'\cup\{m\}, i} \, \xi_m \Big)^2.
\end{align}
In the third line, the $I'$ indexes all pair sets of cardinality $\floor{N/2} - 1$.\par
The way to get to the desired upper bound of this expression is to insert the indicator functions $\mathds1_{m\notin I'}$ and  $\mathds1_{i\notin\phi(I'\cup\{m\})}$ into the final line of eq.~\ref{eq:ex_val_inner_prod}. While these indicator functions are already incorporated in the definition of the coefficients $A_{I, i}$ and hence may seem redundant, they allow for clever use of the Cauchy-Schwarz inequality in two different ways, which leads to a system of inequalities from which we can obtain the upper bound. The first of these is nearly identical to that presented by Tennie et al \cite[appendix A, eq. A3]{Tennie2017}, and is as follows,
\begin{align}\label{eq:ex_val_ineq1}
    \bra\Psi\b_{\vcg\xi}\t\b_{\vcg\xi}\ket\Psi &= \sum_{I', i} \Big( \sum_m A_{I'\cup\{m\}, i} \, \xi_m \mathds1_{m\notin I'}\Big)^2 \nonumber\\
    &\leq \sum_{I', i} \sum_l A_{I'\cup\{l\}, i}^2 \sum_k (\xi_k \mathds1_{k\notin I'})^2 \nonumber\\
    &= \sum_{I', i} \sum_l A_{I'\cup\{l\}, i}^2 \sum_{k\notin I'} \xi_k^2 \nonumber\\
    &= \sum_{I, i} \sum_{l\in I} A_{I, i}^2 \sum_{k\notin I\setminus\{l\}} \xi_k^2 \nonumber\\
    &= \sum_{I, i} A_{I, i}^2 \sum_{l\in I} \Big( \xi_l^2 + \sum_{k\notin I} \xi_k^2 \Big) \nonumber\\
    &= \sum_{I, i} A_{I, i}^2 \Big( \floor{N/2} \sum_{k\notin I} \xi_k^2 + \sum_{l\in I} \xi_l^2 \Big) \nonumber\\
    &= \sum_{I, i} A_{I, i}^2 \Big( \floor{N/2} \sum_{k\notin I} \xi_k^2 + 1 - \sum_{l\notin I} \xi_l^2 \Big) \nonumber\\
    &= 1 + (\floor{N/2} - 1) \sum_{I, i} A_{I, i}^2 \sum_{k\notin I} \xi_k^2.
\end{align}
In the penultimate line, we used the normalisation of $\vcg\xi$, and in the last line that of $\ket\Psi$.\par
For the second way, it becomes important that the unpaired fermion takes away a site pair that could otherwise be occupied by a pair of fermions:
\begin{align}\label{eq:ex_val_ineq2}
    \bra\Psi\b_{\vcg\xi}\t\b_{\vcg\xi}\ket\Psi &= \sum_{I', i} \Big( \sum_m A_{I'\cup\{m\}, i} \, \xi_m \mathds1_{m\notin I'} \mathds1_{i\notin\phi(I'\cup\{m\})} \Big)^2 \nonumber\\
    &\leq \sum_{I', i} \sum_l A_{I'\cup\{l\}, i}^2 \, \xi_l^2 \sum_k \mathds1_{k\notin I'} \mathds1_{i\notin\phi(I'\cup\{k\})} \nonumber\\
    &= \sum_{I', i} \sum_l A_{I'\cup\{l\}, i}^2 \, \xi_l^2 \, (\floor{\L/2} - \floor{N/2}) \nonumber\\
    &= (\floor{\L/2} - \floor{N/2}) \sum_{I, i} \sum_{l\in I} A_{I, i}^2 \, \xi_l^2 \nonumber\\
    &= (\floor{\L/2} - \floor{N/2}) \Big( 1 - \sum_{I, i} \sum_{l\notin I} A_{I, i}^2 \, \xi_l^2 \Big).
\end{align}
Again, in the last line we used the normalisation of $\vcg\xi$ and $\ket\Psi$.\par
When we take the appropriate linear combination of inequalities \ref{eq:ex_val_ineq1} and \ref{eq:ex_val_ineq2}, the sum $\sum_{I, i} A_{I, i}^2 \sum_{k\notin I} \xi_k^2$ cancels out,
\begin{align}
    & \frac{\floor{\L/2} - \floor{N/2}}{\floor{N/2} - 1}\bra\Psi\b_{\vcg\xi}\t\b_{\vcg\xi}\ket\Psi + \bra\Psi\b_{\vcg\xi}\t\b_{\vcg\xi}\ket\Psi \nonumber\\
    &~~~~~~ \leq (\floor{\L/2} - \floor{N/2}) \Big( 1 + \frac1{N-1} \Big),
\end{align}
and we find
\begin{align}
    \bra\Psi\b_{\vcg\xi}\t\b_{\vcg\xi}\ket\Psi \leq \frac{\floor{N/2}(\floor{\L/2} - \floor{N/2})}{\floor{\L/2} - 1}
\end{align}
as desired.

\clearpage
\onecolumngrid

\section{Proof of theorem \ref{prop:even_odd_bounds_tight}}\label{app:proof_prop_even_odd_bounds_tight}
In the following, let $\ell = \floor{\L/2}$. For the even case, when we take $\xi_m = 1/\sqrt\ell \, \forall m$ and $\ket\Psi$ the maximally symmetric state
\begin{align}
    \ket\Psi = \binom{\ell}{N/2}^{-1/2} \sum_{\P:|\P|=N/2} \ket{\P, \emptyset}
\end{align}
where $\ket{\P, \emptyset} = \prod_{k\in\P} \b_k\t\ket{\rm vac}$, then a straightforward calculation shows that
\begin{align}
    \bra\Psi \b_{\vcg\xi}\t\b_{\vcg\xi}\ket\Psi 
    &= \frac1\ell \binom\ell{N/2}^{-1} \sum_{\P, \P'} \sum_{mn} \bra{\rm vac} \Big( \prod_{k\in \P} \b_k \Big) \b_m\t \b_n \Big( \prod_{l\in \P'} \b_l\t \Big) \ket{\rm vac} \nonumber\\[2mm]
    &= \frac1\ell \binom\ell{N/2}^{-1} \sum_{\P, \P'} \sum_{\substack{m\in \P \\ n\in \P'}} \delta_{\P\setminus\{m\}, \P'\setminus\{n\}} \nonumber\\
    &= \frac1\ell \binom\ell{N/2}^{-1} \binom\ell{N/2} (N/2)(\ell - N/2 + 1) \nonumber\\[2mm]
    &= \frac{N/2}{\floor{\L/2}}(\floor{\L/2} - N/2 + 1). \label{eq:even_verzadigd}
\end{align}
In the odd case then, where $\xi_m = 1/\sqrt{\ell - 1} \; \forall m\neq \ell$, $\xi_\ell = 0$ and
\begin{align}
    \ket\Psi = \binom{\ell - 1}{\floor{N/2}}^{-1/2} \sum_{\substack{\P:|\P|=\floor{N/2} \\ \ell\notin \P}} \ket{\P, \{2\ell - 1\}}
\end{align}
where $\ket{\P, \{i\}} = \tilde\a_i\t\prod_{k\in\P} \b_k\t\ket{\rm vac}$, we find
\begin{align}
    \bra\Psi \b_{\vcg\xi}\t\b_{\vcg\xi}\ket\Psi &= \frac1{\ell - 1} \binom{\ell - 1}{\floor{N/2}}^{-1} \!\!\!\! \sum_{\substack{\P, \P' \\ \ell\notin \P, \P'}} \sum_{m, n\neq \ell} \bra{\rm vac} \Big( \prod_{k\in \P} \b_k \Big) \tilde\a_{2\ell - 1} \b_m\t \b_n \tilde\a_{2\ell - 1}\t \Big( \prod_{l\in \P'} \b_l\t \Big) \ket{\rm vac} \nonumber\\[4mm]
    &= \frac1{\ell - 1} \binom{\ell - 1}{\floor{N/2}}^{-1} \sum_{\substack{\P, \P' \\ \ell\notin \P, \P'}} \sum_{m, n\neq \ell} \bra{\rm vac} \Big( \prod_{k\in \P} \b_k \Big) 
    ({\underbrace{\delta_{m\ell} \tilde\a_{2m}\t}_{=\;0}} + \b_m\t \tilde\a_{2\ell - 1}) ({\underbrace{\delta_{n\ell}\tilde\a_{2n}}_{=\;0}} + \tilde\a_{2\ell - 1}\t \b_n)
    \Big( \prod_{l\in \P'} \b_l\t \Big) \ket{\rm vac} \nonumber\\[4mm]
    &= \frac1{\ell - 1} \binom{\ell - 1}{\floor{N/2}}^{-1} \!\!\!\! \sum_{\substack{\P, \P' \\ \ell\notin \P, \P'}} \sum_{m, n\neq \ell} \bra{\rm vac} \Big( \prod_{k\in \P} \b_k \Big) \b_m\t \b_n \Big( \prod_{l\in \P'} \b_l\t \Big) \ket{\rm vac} \nonumber\\[2mm]
    &= \frac{N/2}{\floor{\L/2} - 1}(\floor{\L/2} - N/2).
\end{align}
In the last line, we directly used the result of eq.~\ref{eq:even_verzadigd}.

\end{document}